\documentclass[journal=ancac3,manuscript=article]{achemso}

\usepackage{chemformula} 
\usepackage{amsmath}
\usepackage{inputenc}
\usepackage{textcomp}
\usepackage{graphicx}
\usepackage{breqn}
 \usepackage{tikz}
\usepackage[T1]{fontenc} 
\usepackage{dcolumn}
\usepackage{bm}
\usepackage{mathptmx}
\usepackage{etoolbox}

\author{Adarsh B Vasista}
\email{a.vasista@exeter.ac.uk}
\affiliation
{Department of Physics and Astronomy, University of Exeter, United Kingdom}
\author{Kishan S Menghrajani}
\affiliation
{Department of Physics and Astronomy, University of Exeter, United Kingdom}
\author{William L Barnes}
\affiliation[Unknown University]
{Department of Physics and Astronomy, University of Exeter, United Kingdom}
\title{Effect of molecular absorption and vibrational modes in polariton assisted photoemission from a layered molecular material}

\begin{document}

\begin{tocentry}
\centering
\includegraphics{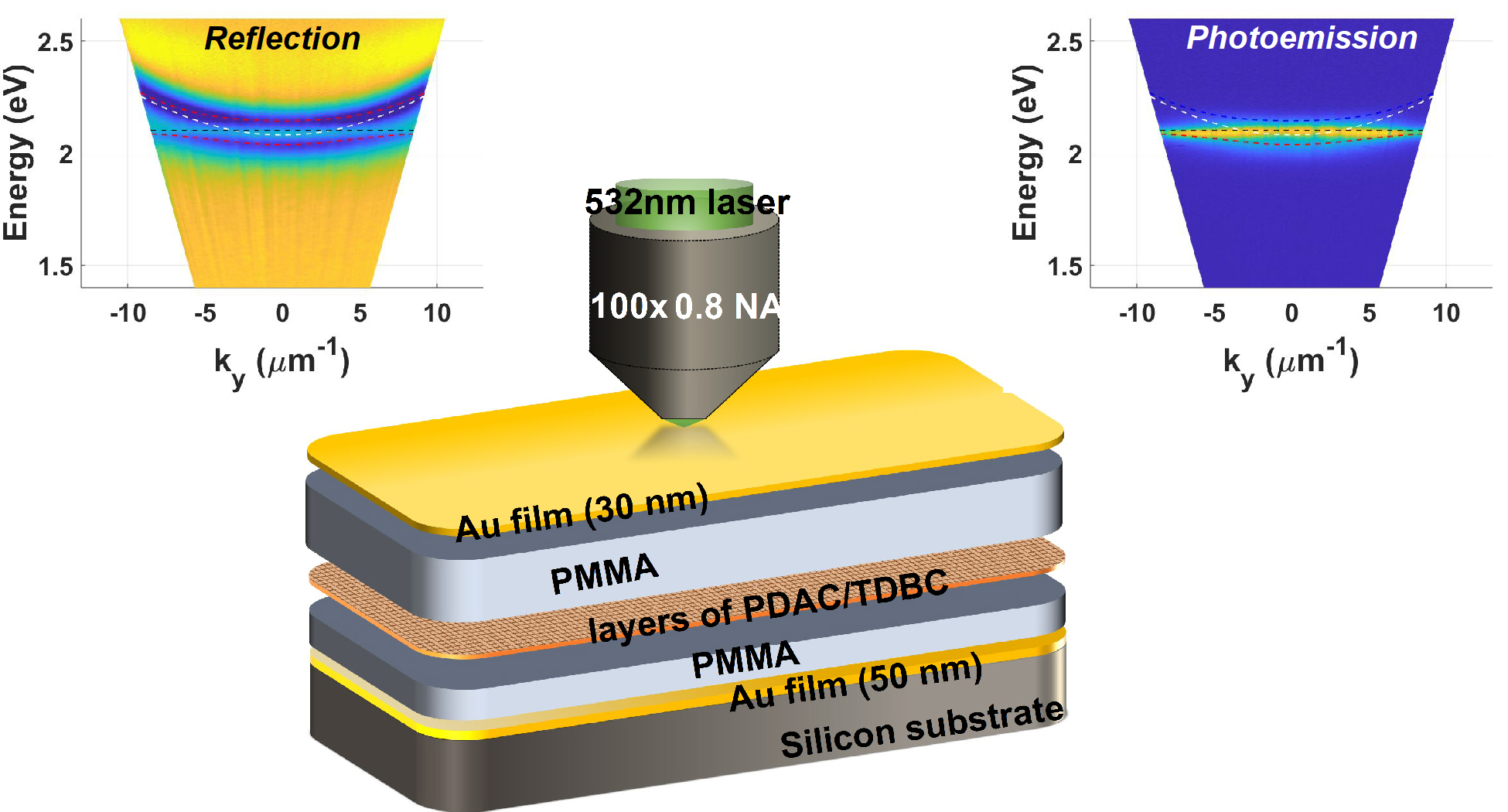}

\end{tocentry}

\begin{abstract}
The way molecules absorb, transfer, and emit light can be modified by coupling them to optical cavities. The extent of the modification is often defined by the cavity-molecule coupling strength, which depends on the number of coupled molecules. We experimentally and numerically study the evolution of photoemission from a thin layered J-aggregated molecular material strongly coupled to a Fabry-Perot microcavity as a function of the number of coupled layers. We unveil an important difference between the strong coupling signatures obtained from reflection spectroscopy and from polariton assisted photoluminescence. We also study the effect of the vibrational modes supported by the molecular material on the polariton assisted emission both for a focused laser beam and for normally incident excitation, for two different excitation wavelengths: a laser in resonance with the lower polariton branch, and a laser not in resonance. We found that the Raman scattered photons play an important role in populating the lower polariton branch, especially when the system was excited with a laser in resonance with the lower polariton branch. We also found that the polariton assisted photoemission depends on the extent of modification of the molecular absorption induced by the molecule-cavity coupling.  
\end{abstract}
\maketitle
\vskip1.0cm
\section{Introduction}
Controlling photoemission from molecular materials has wide implications in designing display devices\cite{1,2,3}, sensors\cite{4,5}, and light emitting diodes\cite{6,7,11}. Various mechanisms have been utilized to modify photophysics of materials for example by doping\cite{8,9} and preparation of heterojunctions\cite{10,12}. Recently optical control of photoluminescence by coupling a molecular material to optical cavities has gained prominence\cite{13}. The extent of the modification of the photophysics then depends upon the coupling strength between the cavity and the material. If the coupling strength between the molecular material and the cavity is less than the losses in the system, then the coupled system is said to be in the weak coupling regime. Here, one can alter the radiative relaxation rate\cite{14} and the directionality of emission\cite{15,37}. If the coupling strength is more than the losses, then we enter into strong coupling regime where one drastically modifies the energy landscape of the molecular material-cavity system.\\

Strong molecule-cavity coupling has been shown to modify various physical processes such as energy transfer\cite{16}, triplet state dynamics\cite{17}, photo-oxidation\cite{18}, and lasing\cite{19}. Due to the drastic modification of molecular energy levels, strong coupling provides a unique way to control photoemission from molecular materials. Various studies have been done to understand photoluminescence (PL) from J-aggregated molecules\cite{20,21,22,24}, layered dichalcogenides\cite{23}, and Carbon nanotubes \cite{42}. However, the relation between molecular absorption and PL under strong coupling regime has not been extensively studied. Also, most of the studies have focused on the behavior of PL produced by non-resonant excitation \cite{20,21,22}, there are very few reports on probing PL by resonantly pumping the lower polariton branch\cite{39,40}. A comprehensive understanding of the evolution of PL as a function excitation mechanism and the modification of molecular absorption seems to be lacking in the literature. This fundamental knowledge gap has wide implications in designing polariton mediated light emitting devices.

Strong light-matter coupling is often probed using reflection, transmission, or scattering spectroscopy.\cite{44,45,46} It has been shown that the signature of strong coupling depends in a critical way on the method of probing the system\cite{25}, and indeed aberrations due to a high numerical aperture may be misrepresented as strong coupling signatures\cite{41}. Hence, it is very important to corroborate the PL from strongly coupled system with  reflection/transmission signatures from the same system. In addition PL is a two step process- absorption followed by relaxation. Hence PL from a coupled system will depend on the extent of modification induced in the absorption spectrum of the material.

With this motivation we study PL signatures from layers of   J-aggregated 5,5',6,6'-tetrachloro-1,1'-diethyl-3,3'-di(4-sulfobutyl)-benzimidazolocarbocyanine (TDBC) dye molecules prepared using a layer-by-layer process coupled to a Fabry-Perot microcavity. Such layered J-aggregates have shown to possess large oscillator strength\cite{26}, high absorption\cite{27}, and high quantum yield\cite{28} making them an ideal system to study photoemission. We systematically study, using a custom built Fourier microscope, the evolution of PL by increasing the number of layers of dye molecules and the cavity detuning. In addition we also study the effect of excitation wavevector and energy on the PL emission from such a layered molecular material. 

We unveil an important difference between absorption of the hybrid microcavity-molecule system and the energy dissipation inside the molecular material, and we discuss the implications for control of molecular PL. We also extend our understanding of the role of the vibrational modes of the molecules in polariton assisted PL by exciting the hybrid system with a laser in resonance with the lower polariton branch. The experimental results were analysed using a simple coupled oscillator model and we performed finite element method (FEM) based numerical modelling to understand the system better.

\section{Results and Discussions}
\begin{figure}[h!]
    \centering
    \includegraphics[width=\linewidth]{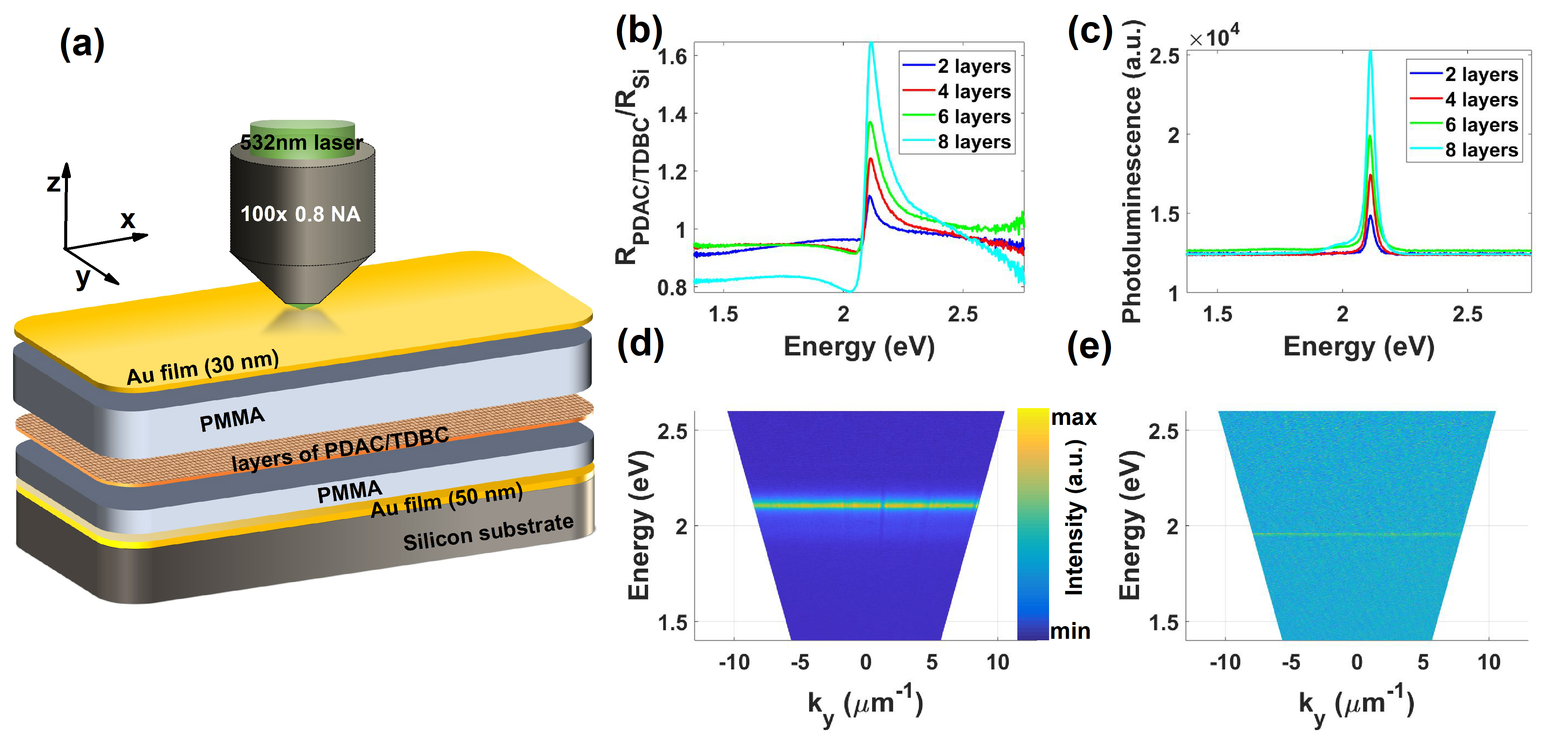}
    \caption{\textit{Schematic of the experiment and spectra of “bare” molecular resonance and photoluminescence.} (a) Schematic representing the system under study. A microcavity was prepared by sandwiching layers of J-aggregated dye (TDBC) between thin gold films and PMMA. The thickness of the bottom PMMA layer was kept at ~100nm and the thickness of the top layer was varied to tune the cavity resonance. An individual microcavity was then probed using a 100x, 0.8NA objective lens. (b) Experimentally measured reflectivity spectra of different layers of PDAC/TDBC on silicon substrate. (c) Experimentally measured photoluminescence spectra of different layers of PDAC/TDBC on silicon substrate. (d) and (e) are angle resolved photoluminescence plots from 8 layers of PDAC/TDBC on a silicon substrate excited with 532 nm and 633 nm lasers respectively. }
    \label{fig:my_label}
\end{figure}

Figure 1 (a) shows a schematic of the system under study. Layers of J-aggregated TDBC were sandwiched between two PMMA coated gold films. The bottom gold mirror was 50 nm thick. The thickness of PMMA substrate was kept at 100 nm so as to maxmimize the molecule-electric field interaction. The thickness of the top PMMA layer was varied so as to tune the cavity resonance, the top gold film was 30 nm thick. This microcavity was probed using a 100x, 0.8 NA objective lens. For reflectivity measurements a white light source was coupled to the objective lens which focused the light onto the sample. The reflected light was then collected using the same objective lens and projected to the Fourier plane\cite{43} and then analyzed for spectral and wavevector information. For PL measurements, a laser source (532 nm and 633 nm) was focused onto the sample and the PL was collected by the same objective lens. The light was then analyzed for wavevector and spectral information by rejecting the laser line using edge filters. To perform PL measurements for a specific set of incident angles, we focused the input laser beam onto the back-aperture of the objective lens. The PL was then collected using the same lens and analyzed (For further details of the experimental setup, please see section S1 of the supplementary information).

To understand the response of the bare molecular film, we performed reflectivity and PL measurements on layers of PDAC/TDBC on a silicon substrate. Figure 1 (b) shows experimentally measured white light reflection spectra as a function of increasing number of molecular layers. We can see that the intensity of reflection near the molecular resonance ($E_{mol}$) increases with the number of layers, as expected. Figure 1 (c) shows corresponding PL spectra collected by exciting the system with a 532 nm CW laser. The intensity of the PL emission of the bare J-aggregated dye increases as a function of the number of PDAC/TDBC layers in keeping with the reflectivity data. Figure 1 (d) and (e) show angle resolved dispersion plots of PL collected from 8 layers of PDAC/TDBC coated on a bare silicon substrate. For 532 nm excitation ($E_{exc}>E_{mol}$),we see angle independent emission around 590 nm, this is J-aggregate PDAC/TDBC PL. To study the response of the system at a lower energy, we excited it with 633 nm laser ($E_{exc}<E_{mol}$) and the result is shown in figure 1 (e). Since the energy of excitation now is lower than the molecular absorption of the J-aggregated dye we see no emission from the system. The faint line around $E=1.96 eV$ is due to the scattered laser line entering the spectrometer.
\subsection{\textit{Effect of number of molecules (N) and detuning ($\Delta$) on polariton assisted PL}}
\begin{figure}[h!]
    \centering
    \includegraphics[width=\linewidth]{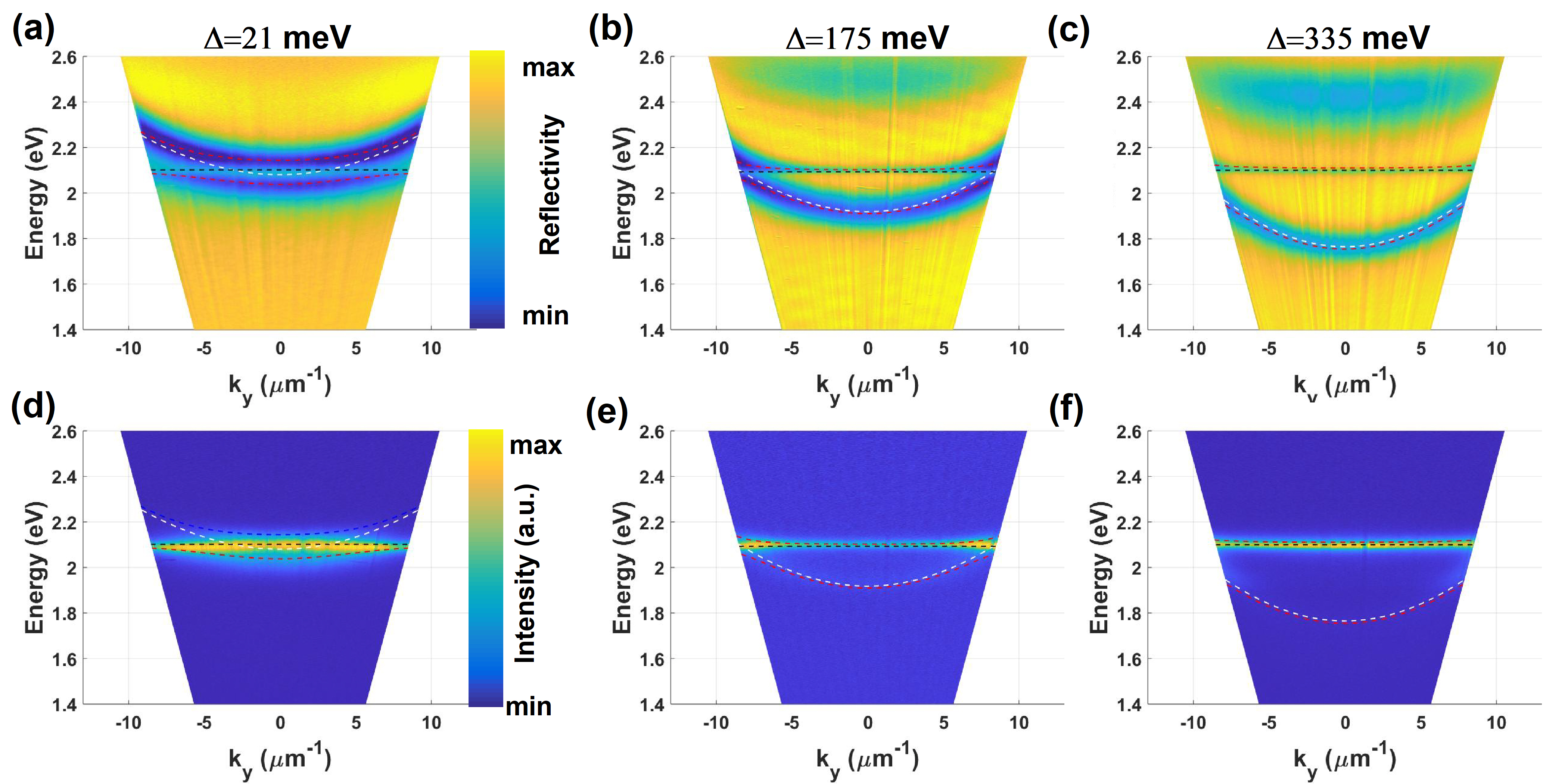}
    \caption{\textit{Polariton assisted Photoluminescence as a function of detuning (2 layers)}. (a)-(c) Experimentally measured angle resolved white light reflection plots for 2 layers of PDAC/TDBC coupled to a microcavity with detunings of 21, 175, and 335 meV respectively. (d)-(e) are the corresponding photoluminescence plots by exciting the system with a focused beam of 532nm laser. The excitation polarization was kept transverse magnetic (TM). The superimposing dashed lines were calculated using a simple coupled oscillator model to estimate the coupling strength. The dashed white line represents the uncoupled cavity mode, the dashed black line represents the molecular absorption, and the dashed red line represents the polaritons.   }
    \label{fig:my_label}
\end{figure}
To start with we studied two layers of PDAC/TDBC coupled to a microcavity. Figure 2 (a)-(c) show the angle resolved white light reflectivity plots for cavity detunings of ($\Delta = E_{mol}-E_{cavity}$) 21 meV, 175 meV, and 335 meV respectively. The input polarization was kept as TM for all the cases (similar results were obtained with TE poalrization, data not shown). We can see that the microcavity mode is split by the molecular resonance into two polariton branches. The extent of the splitting was estimated by fitting the experimentally measured data with a simple coupled oscillator model (see numerical analysis and modelling section). The result of a simple coupled oscillator fit is superimposed on the experimental data (see figures 2 (a)-(c)). The estimated coupling strength, $2g$, was found to be 104$\pm$1 meV, 80 $\pm$2 meV, and 100 $\pm$2 meV respectively for detunings of 21 meV, 175 meV, and 335 meV respectively. This slight variation in the coupling strength can be attributed to the inhomogeneity of the molecular adsorption and variations of the lower PMMA layer thickness. However, in all three cases we see that the value of coupling strength, $2g$, is greater than the mean of linewidths of the cavity resonance ($\gamma_{cavity} = 70 meV$) and the molecular resonance ($\gamma_{mol}=53 meV$) showing that the cavities were in the strong coupling regime\cite{29,30}. The mixing fractions calculated using Hopfield analysis are shown in section S2 of the supplementary information.

Individual cavities were excited with a focused beam of 532 nm laser to study the PL emission. Figure 2 (d)-(f) show the angle resolved PL plots from the microcavities. In all of the plots one can notice an angle independent strong emission around 2.1 eV, this is the bare molecular emission maximum. These plots indicate that very little modification in the molecular PL arises from coupling two layers of PDAC/TDBC to the microcavity. Note that the white light reflectivity shows a clear signature of strong coupling.

\begin{figure}[h!]
    \centering
    \includegraphics[width=\linewidth]{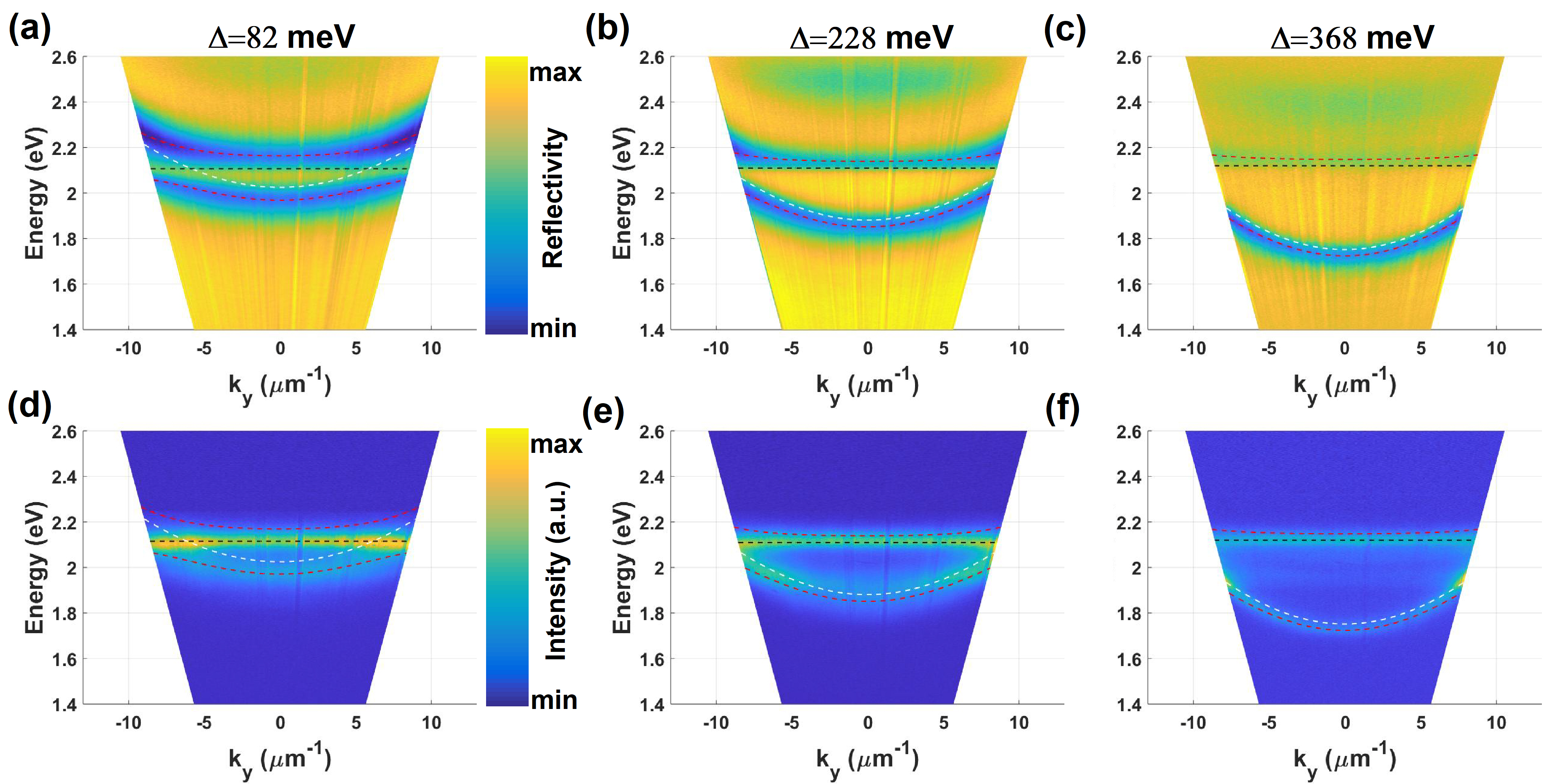}
    \caption{\textit{Polariton mediated Photoluminescence as a function of detuning (8 layers)}. (a)-(c) Experimentally measured angle resolved white light reflection plots for 8 layers of PDAC/TDBC coupled to a microcavity with detunings of 82, 228, and 368 meV respectively. (d)-(e) are the corresponding photoluminescence plots by exciting the system with 532nm laser. The excitation polarization was kept transverse magnetic (TM). The superimposing dashed lines were calculated using a simple coupled oscillator model to estimate the coupling strength. The dashed white line represents the uncoupled cavity mode, the dashed black line represents the molecular absorption, and the dashed red line represents the polaritons.   }
    \label{fig:my_label}
\end{figure}
To further understand molecular emission from layered organic molecules, we coupled eight layers of PDAC/TDBC to a microcavity. Figure 3 (a)-(c) show angle resolved white light reflectivity plots for 8 layers of PDAC/TDBC coupled to a microcavity with $\Delta$= 82 meV, 228 meV, and 368 meV respectively. The extent of the splitting has considerably increased by increasing the number of molecular layers, as expected. We fit the experimental data using a simple coupled oscillator model to estimate the coupling strength. The estimated value of $2g$ was 177$\pm$1 meV, 175$\pm$2 meV, and 210$\pm$5 meV respectively for microcavity with $\Delta$= 82 meV, 228 meV, and 371 meV. Since the coupling strength, $2g$, is greater than the mean of cavity and molecular absorption linewidths, we can again say that the system is in strong coupling regime. 

We excited the microcavities with a focused beam of a 532 nm laser and collected the PL. Figure 3 (d)-(f) show angle resolved PL plots for different cavity detunings. In all three cases, we see angle independent emission at around 2.1 eV corresponding to \textit{uncoupled} molecules and an additional angle dependent emission at lower energies due to the lower polariton assisted emission, in contrast with the two layer PDAC/TDBC case. However, we do not see any upper polariton assisted emission, which can be attributed to the very fast relaxation times associated with the upper polariton branch\cite{31,32,24}. Polariton assisted PL in the case of TDBC molecules has been studied in the past, including the details of possible processes involved\cite{21}. In the case of TDBC molecules, which show a negligible Stokes shift in the PL, lower polariton assisted emission is mainly due to vibrationally assisted scattering\cite{33,21}. What this means is that there is a  transfer of energy from reservoir states to the lower polariton branch which is maximal when there exists a vibrational energy quantum which facilitates the scattering. We discuss these processes in detail later in the article.

We systematically studied the evolution of the molecule-cavity coupling and the resulting molecular PL by increasing the number of coupled molecular layers. Section S3 of supplementary information shows the white light reflectivity and angle resolved PL from the coupled system. As one increases the number of molecular layers, we see the emergence of lower polariton assisted PL and at the same time a decrease of the intensity of the PL due to \textit{uncoupled} excitons.

\subsection{\textit{Numerical modelling of molecule-cavity coupling}}
\begin{figure}[h!]
    \centering
    \includegraphics[width=\linewidth]{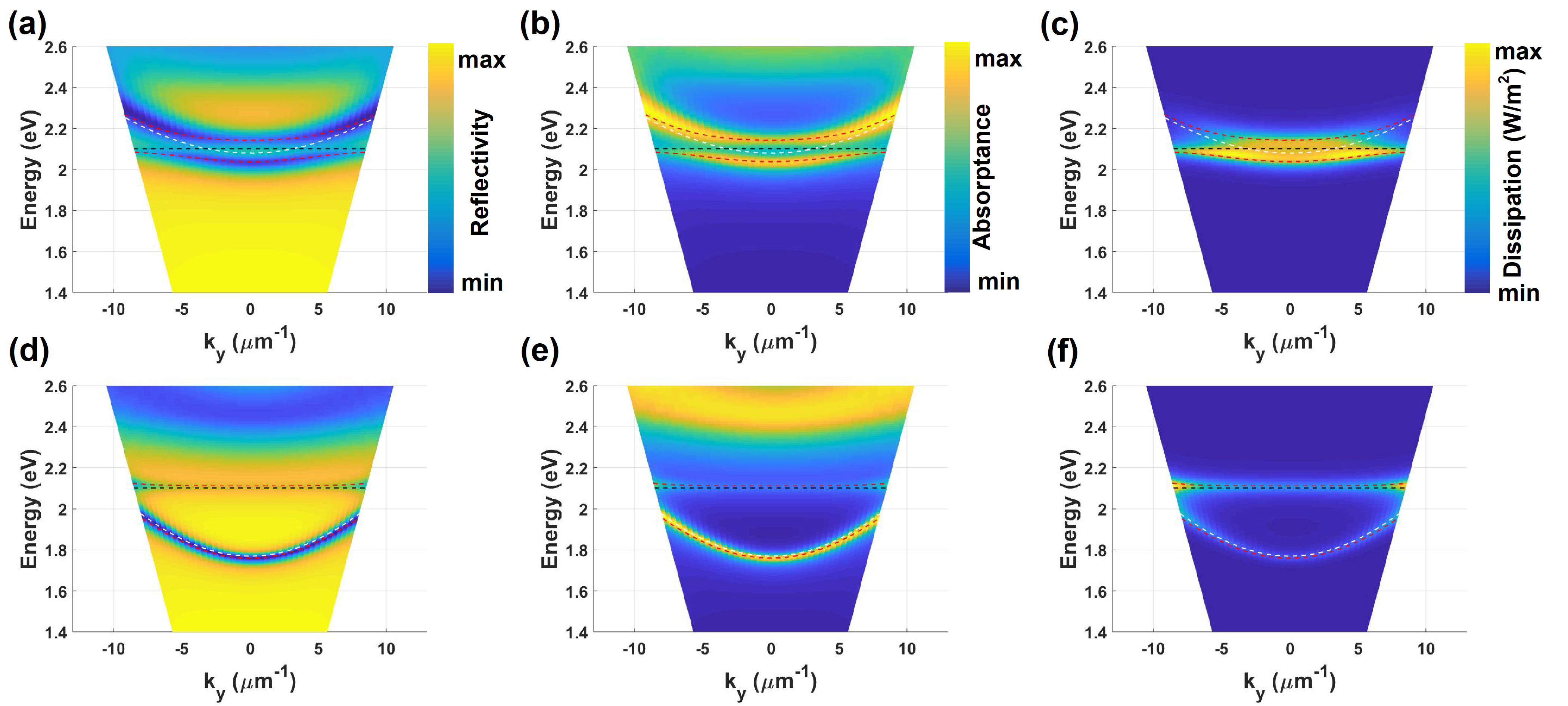}
    \caption{\textit{Numerical modelling of molecule-cavity coupling (2 layers)}. (a) Numerically calculated angle resolved reflectivity spectra of a microcavity coupled to 2 layers of PDAC/TDBC with cavity detuning of 21 meV. (b) Corresponding numerically calculated angle resolved absorption spectra. (c) Numerically calculated angle resolved dissipation density inside the PDAC/TDBC layer. (d) Numerically calculated angle resolved reflectivity spectra of a microcavity coupled to 2 layers of PDAC/TDBC with cavity detuning of 335 meV. (b) Corresponding numerically calculated angle resolved absorption spectra. (c) Numerically calculated angle resolved dissipation density inside the PDAC/TDBC layer. The superimposing dashed lines were calculated using a simple coupled oscillator model to estimate the coupling strength. The dashed white line represents the uncoupled cavity mode, the dashed black line represents the molecular absorption, and the dashed red line represents the polaritons.}
    \label{fig:my_label}
\end{figure}

We performed finite element method (FEM) based numerical calculations using COMSOL multiphysics to understand the molecule-cavity coupling in layered PDAC/TDBC (see section on numerical modelling for further details). The optical properties of the PDAC/TDBC layer were modelled using a uniaxial lorentzian oscillator. Figure 4 (a) shows numerically calculated angle resolved reflectivity spectra for 2 layers of PDAC/TDBC coupled to a microcavity, nicely reproducing the experimentally measured white light reflectivity data of figure 2 (a). We clearly see splitting and anticrossing of the cavity mode showing that the system is in the strong coupling regime. Further, we calculated the absorption spectra of the system (\textit{1-reflection-transmission}), shown in figure 4 (b). We see clear splitting and anticrossing of the cavity mode resulting in polaritons in the absorption spectra as well. Even with clear splitting and anti-crossing in the absorption and the reflection spectra there is no evidence of polariton assisted PL emission (see figure 2 (d) -(f)). To understand this difference between the PL and reflection/absorption spectra, we numerically integrated the electromagnetic dissipation density inside the PDAC/TDBC layer to give an estimate of the absorption of just the molecules. Figure 4 (c) shows the angle resolved dissipation inside the PDAC/TDBC layer. We can see clear difference between the absorption of the total system (microcavity+molecular layer) and just the PDAC/TDBC molecular layer. The absorption in just the molecular layer has not changed as much as the absorption or reflectivty data of the whole system portrays. Hence, once the microcavity containing 2 layers of PDAC/TDBC is excited using a 532 nm laser, due to a slight modification of the  molecular absorption, the majority of resulting PL was energetically unmodified compared to that from the bare molecules.

To study the effect of detuning on the molecular absorption, we repeated the same calculation for the cavity detuning of 335 meV. Figure 4 (d) and (e) show the angle resolved reflectivity and absorptance of the microcavity+molecule system resembling the experimental data (see figure 2). The absorption and the reflectivity spectra show clear splitting of the cavity mode resulting from the formation of polaritons. However, the electromagnetic dissipation calculated inside the molecular layer shows very minimal perturbation to the absorption of the molecule, and this is reflected in the PL signature. 
\begin{figure}[h!]
    \centering
    \includegraphics[width=\linewidth]{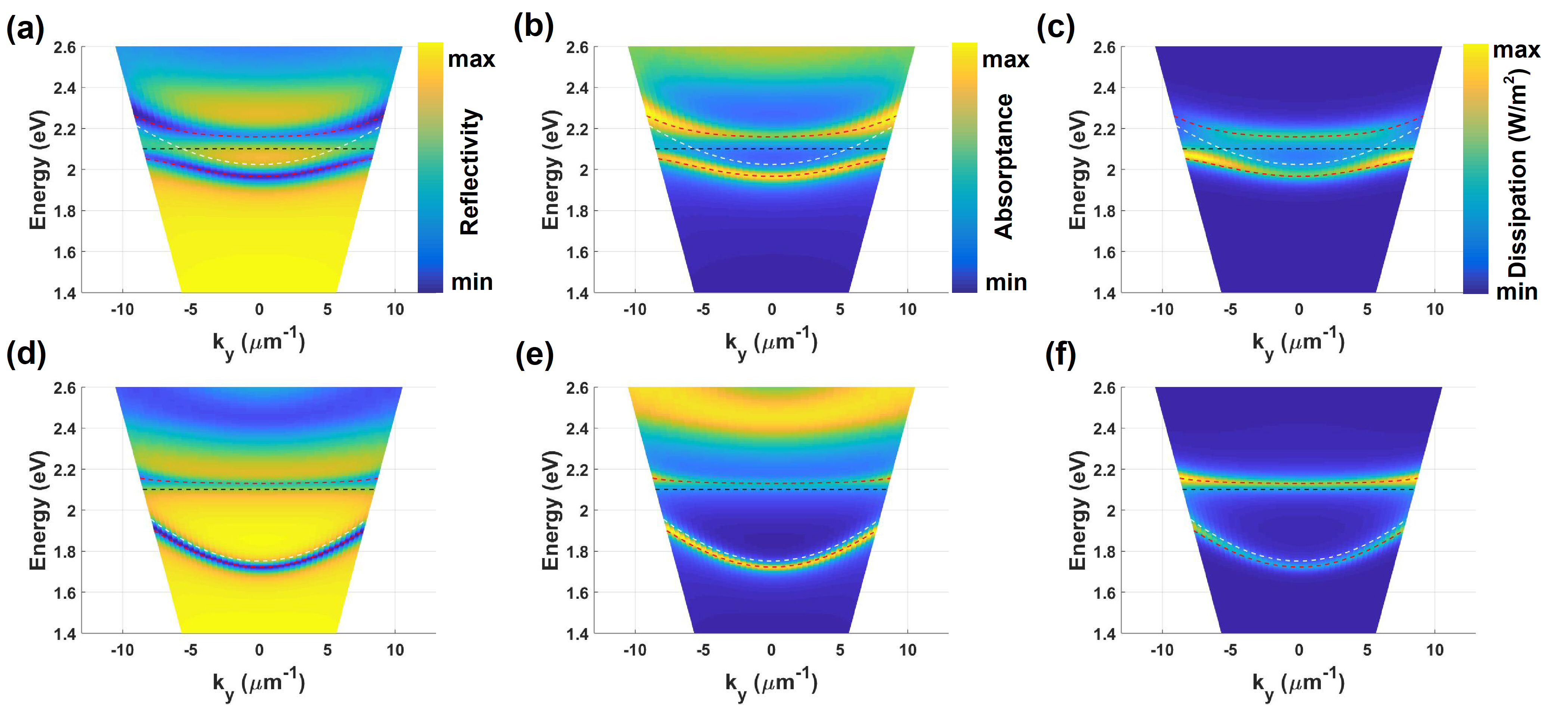}
    \caption{\textit{Numerical modelling of molecule-cavity coupling (8 layers)}. (a) Numerically calculated angle resolved reflectivity spectra of a microcavity coupled to 8 layers of PDAC/TDBC with cavity detuning of 82 meV. (b) Corresponding numerically calculated angle resolved absorption spectra. (c) Numerically calculated angle resolved dissipation density inside the PDAC/TDBC layer. (d) Numerically calculated angle resolved reflectivity spectra of a microcavity coupled to 8 layers of PDAC/TDBC with cavity detuning of 368 meV. (b) Corresponding numerically calculated angle resolved absorption spectra. (c) Numerically calculated angle resolved dissipation density inside the PDAC/TDBC layer. The superimposing dashed lines were calculated using a simple coupled oscillator model to estimate the coupling strength. The dashed white line represents the uncoupled cavity mode, the dashed black line represents the molecular absorption, and the dashed red line represents the polaritons.}
    \label{fig:my_label}
\end{figure}

Figure 5 (a) and (b) show numerically calculated angle resolved reflectivity and absorption for 8 layers of PDAC/TDBC coupled to a microcavity, with a cavity detuning of 82 meV. Now we clearly see splitting and anticrossing of the cavity mode mimicking the experimental data (see figure 3). Figure 5 (c) shows the numerically integrated dissipation density inside the PDAC/TDBC layer clearly indicating the splitting associated with the molecular absorption. This absorption splitting is an indication of the modification of energy landscape of the molecule which is then reflected in the photoluminescence signature. We repeated these calculations for cavity detuning of 368 meV and the results are shown in figure 5 (d)-(f). We can see that the modification of the absorption of the molecule here as well, in contrast to the 2 layer PDAC/TDBC case.  

Figures 4 and 5 shed light on understanding the evolution of molecular PL as a function of the number of layers as well as the detuning. For far-red cavity detuning ($\Delta$> 300 meV) we see that the Hopfield coeffecients of the lower polariton are dominated by the cavity contribution (see section S2 and S3 of supplementary information). It is tempting to assume that, because of almost zero exciton contribution to the lower polariton branch there would be no or minimal photoluminescence. However, this is not the case as shown for 8 layers of PDAC/TDBC coupled to the microcavity (see figure 3 (f)). The modification of energy landscape, through splitting of the molecular absorption, is the key to generate polariton assisted molecular PL. However, it is also important to note that the cavity-molecule mixing fractions play a critical role in polariton scattering thus affecting the polariton relaxation associated with the bottleneck effect\cite{20}. 

These results indicate that an apparent split and anti-crossing observed in reflection may not result in a significant modification of the molecular absorption which is needed to alter the photophysics of molecules. 

\subsection{\textit{Effect of excitation mechanism}}
To understand the processes involved in polariton assisted PL in more detail we excited a microcavity with 8 layers of PDAC/TDBC with a 532 nm excitation - to pump the system at an energy above the upper polariton branch, and with 633 nm - to excite only the lower polariton branch. Figure 6 shows the measured angle resolved PL spectra by altering the excitation mechanism. We chose a microcavity with a detuning of 368 meV so that the LP could be excited using 633 nm laser.     
\begin{figure}[h!]
    \centering
    \includegraphics[width=\linewidth]{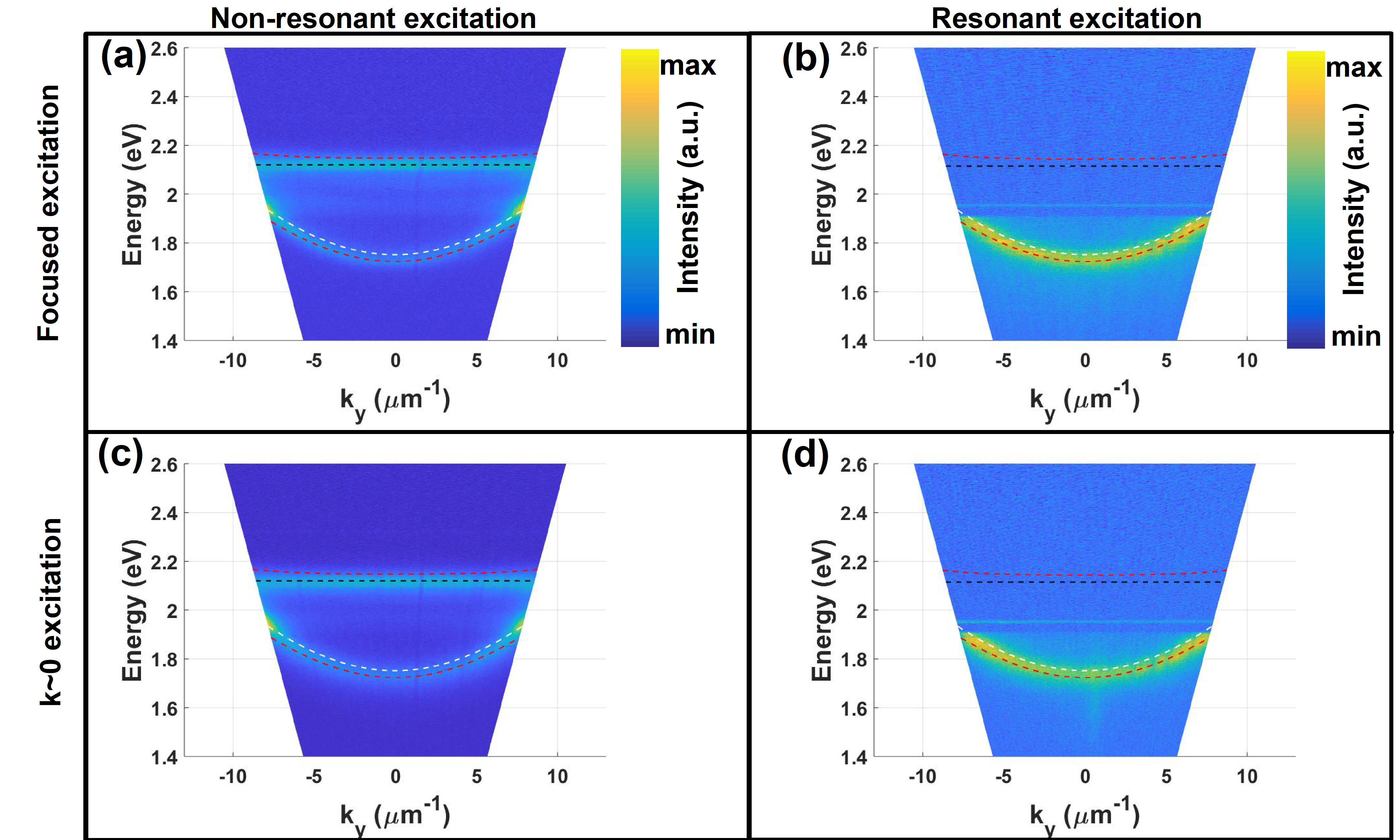}
    \caption{\textit{Polariton assisted Photoluminescence as a function of excitation mechanism (8 layers)}. (a) and (b) are experimentally measured angle resolved PL spectra for 8 layers of PDAC/TDBC coupled to a microcavity excited using a focused beam of 532 nm and 633 nm laser respectively (c) and (d) are experimentally measured angle resolved PL spectra for 8 layers of PDAC/TDBC coupled to a microcavity excited with a laser beam  at k$\sim$0 of 532 nm and 633 nm laser respectively. The superimposing dashed lines were calculated using a simple coupled oscillator model to estimate the coupling strength. The dashed white line represents the uncoupled cavity mode, the black line represents the molecular absorption, and the dashed red line represents the polaritons.}
    \label{fig:my_label}
\end{figure}


Figure 6 (a) shows angle resolved PL spectra collected by exciting a microcavity with 8 layers of PDAC/TDBC using a focused beam from a 532 nm laser. For the non-resonant excitation case (532 nm laser) we see that the LP assisted emission shows intense emission at large wavevectors around 1.95 eV. The vibrational modes associated with the molecule assist in the relaxation of the reservoir states, thus selectively populating the LPB at specific energies,when the vibrational quanta supported by the molecule match the energy difference between the reservoir and the LPB. The vibrational modes of PDAC/TDBC overlap with the LPB around 1.95 eV. For this reason, there is a non-uniform intensity distribution along the LPB (see section S4 of the supplementary information for more details on vibrational modes and its effect on LPB emission). The more interesting case is when the system is excited using a laser resonant with the LPB (633 nm). Figure 6 (b) shows the angle resolved PL when the system was excited using a focused beam of 633 nm laser. In this case, we suggest that the observed molecular PL is produced by two mechanisms: 
\begin{itemize}
    \item The relaxation of LPB which was directly excited using the 633 nm laser. This is possible because of (a) modified absorption of the molecule resulting in a low energy absorption band due to the coupling with the microcavity (see figure 5 (f)) (b) the input laser beam has a wide range of wavevectors. This allows us to populate the LPB with the 633 nm laser, which eventually relaxes through polariton scattering. 
    
    \item The Incident photons that are Raman scattered by the molecule, populating the LPB which eventually undergoes relaxation. We clearly see Raman peaks of PDAC/TDBC overlying the broad molecular PL (see section S4 of supplementary information) confirming this mode of excitation/relaxation.
\end{itemize}
 
To better understand the role of vibrational quanta in the polariton assisted PL we excited the microcavity with input wavevector k$\sim$0. To achieve this we focus the input laser beam to the back-aperture of the objective lens\cite{38}. Figure 6 (c) shows the experimentally measured angle resolved PL spectrum when the microcavity was excited using a 532 nm laser beam. We see no recognizable difference between the focused excitation case and the k$\sim$0 case. We suggest that this is because the reservoir states play a critical role in providing all possible wavevectors needed to populate the LPB. Figure 6 (d) shows angle resolved PL of the microcavity by exciting it with a 633 nm laser at k$\sim$0. An important aspect to note here is that we do not have access to reservoir states of the molecules and the excitation mechanism is such that the system doesn't have access to an excited eigen state (no energy eigen state at k$\sim$0 around excitation energy, 1.959 eV). The mechanism involved in populating the LPB in this case we suggest is predominantly through Raman scattering from the molecules. The secondary Raman scattered photons have a sufficiently wide range of momentum vectors to populate the lower polariton branch, the LPB then relaxes through polariton scattering. Hence even though we excite the microcavity with a very small range of wavevectors, we see emission from the whole of the lower polariton branch as shown in figure 6 (d). We see sharp Raman peaks overlying the PL spectra which confirms this hypothesis (see section S4 of the supplementary information).

\section{Conclusions}
To summarize, we systematically studied the molecular PL from few layers of PDAC/TDBC molecules coupled to a microcavity as a function of cavity detuning ($\Delta$),the  number of coupled molecules ($N$), and the excitation mechanism. We highlight the difference between strong coupling signatures seen in reflection spectroscopy and molecular PL and explain the apparent difference using numerical calculations. We also show the importance of vibrational modes of the molecule in PL, particularly when the system is resonantly excited to the lower polariton branch. We anticipate that these results will have wide implications in understanding and designing polariton assisted light emitting devices.

\section*{Sample preparation}
The bottom mirror was prepared by thermally evaporating 50 nm of gold  over a silicon substrate and a 100 nm layer of PMMA was added by spin coating to create the substrate layer. Then the molecular film was deposited over the PMMA using a layer-by-layer approach\cite{27}. Briefly, we used a cationic poly(diallyldimethylammonium chloride) (PDAC) solution as the polyelectrolyte binder for anionic TDBC J - aggregate solution. A typical deposition step consists of subsequent dipping the substrate inside a beaker of PDAC solution (20$\%$ by weight in water - diluted 1:1000) and TDBC solution in water (0.01M  diluted 1:10) for 15 minutes each. The substrate was washed with DI water after each immersion and same steps were repeated to deposit multiple layers of PDAC/TDBC. To increase the adhesion we first coat one layer of anionic polystyrene sulfonate (PSS) using the above mentioned process and continue with PDAC - TDBC. Finally the TDBC layer was protected by depositing a layer of PDAC molecules. The superstrate was then prepared by spinning a layer of PMMA (thickness was varied so as to yield different detunings) and finally the top mirror was prepared by thermally evaporating 30 nm of gold. 

\section*{Numerical Analysis and modelling}
The experimental data were analysed and fit using a simple coupled oscillator model given as
\begin{equation}
    \begin{pmatrix}
    E_{cavity}-i\frac{\gamma_{cavity}}{2} & 0\\ 0 & E_{PDAC/TDBC}-i\frac{\gamma_{PDAC/TDBC}}{2}
    \end{pmatrix}
      \begin{pmatrix}
   \alpha \\ \beta
    \end{pmatrix}
    = E_{pol}
    \begin{pmatrix}
       \alpha \\ \beta
    \end{pmatrix}
    \end{equation}
where $E_{cavity}$ is the cavity resonance, $\gamma_{cavity}$ is the cavity resonance linewidth, $E_{PDAC/TDBC}$ is the molecular absorption maximum, $\gamma_{PDAC/TDBC}$ is the molecular absorption linewidth, and $E_{pol}$ gives the energies of the polaritons. The eigen vectors of equation (1) provides the information on mixing fraction, also called Hopfield co-effecients.

Microcavities coupled to layers of PDAC/TDBC were modelled as a z axis oriented stack (see figure 1 (a)) using 2D finite element method (FEM) based numerical simulations. The complex refractive indices of gold, PMMA, and silicon were taken from \cite{34},\cite{35}, and \cite{36} respectively. The permittivity of the PDAC/TDBC system was taken to be uniaxial $\begin{pmatrix}
\epsilon^{x} & 0 & 0\\ 0 & \epsilon^{y} & 0 \\ 0 & 0 & 1.9
\end{pmatrix}$
 with individual terms given by
\begin{equation}
    \epsilon^{x,y}_{PDAC/TDBC}(E)=\epsilon_{Inf}+ \frac{fE_{TDBC}^2}{E_{TDBC}^2-E^2-iE\gamma_{TDBC}}
\end{equation}
where $\epsilon_{Inf}$ is the background permittivity set to 1.9, $E_{TDBC}$ is the molecular absorption maximum set to 2.10143 eV, $f$ is the reduced oscillator strength varied to fit the experimental data, and $\gamma$ is the absorption linewidth set to 53 meV. The input plane wave was set to be TM polarized. The substrate PMMA thicknes was taken to be 100 nm and the superstrate thickness was varied to fit the experimental data for various detunings. The bottom and top gold mirrors were set to have thicknesses of 50 nm and 30 nm respectively. The thickness of an individual PDAC/TDBC layer was taken as 2 nm \cite{27}. 

\section*{Acknowledgement}

The authors thank Wai Jue Tan for his help in preparing samples. The authors acknowledge the support of European Research Council through the Photmat project 
(ERC-2016-AdG-742222
:www.photmat.eu) and the support of The Leverhulme Trust. ABV thanks Sunny Tiwari and Deepak Kumar Sharma for fruitful discussions.

\section*{Data availability}

Research data are available from the University of Exeter repository at https://doi.org/xxxx

\bibliography{main}


\end{document}